\documentclass[twocolumn,aps,showpacs,floatfix,prc]{revtex4}
\usepackage[dvips]{epsfig}
\begin{document}

\title{ Photonuclear reactions of actinides in the giant dipole resonance region }
\author{ Tapan Mukhopadhyay$^1$ and D.N. Basu$^2$ }

\affiliation{Variable  Energy  Cyclotron  Centre, 1/AF Bidhan Nagar, Kolkata 700 064, India }
\email[E-mail 1: ]{tkm@veccal.ernet.in}
\email[E-mail 2: ]{dnb@veccal.ernet.in}

\date{\today }

\begin{abstract}

    Photonuclear reactions at energies covering the giant dipole resonance (GDR) region are analyzed with an approach based on nuclear photoabsorption followed by the process of competition between light particle evaporation and fission for the excited nucleus. The photoabsorption cross section at energies covering the GDR region is contributed by both the Lorentz type GDR cross section and the quasideuteron cross section. The evaporation-fission process of the compound nucleus is simulated in a Monte-Carlo framework. Photofission reaction cross sections are analyzed in a systematic manner in the energy range of $\sim$ 10-20 MeV for the actinides $^{232}$Th, $^{238}$U and $^{237}$Np. Photonuclear cross sections for the medium-mass nuclei $^{63}$Cu and $^{64}$Zn, for which there are no fission events, are also presented. The study reproduces satisfactorily the available experimental data of photofission cross sections at GDR energy region and the increasing trend of nuclear fissility with the fissility parameter $Z^2/A$ for the actinides.
\end{abstract}

\maketitle

\noindent
\section{Introduction}
\label{section1}

    In recent years the study of photonuclear reactions has attracted considerable interest. Availability of intense neutron rich ion beams \cite{Di99} will open new vistas in the study of nuclei very far away from the valley of stability especially in the vicinity of $^{78}$Ni. The use of the energetic electrons is a promising tool to get intense neutron-rich ion beams. Photofission of Uranium is a very powerful mechanism to produce such radioactive beams (RIB). Although the photofission cross section at giant dipole resonance (GDR) energy for $^{238}$U is about an order of magnitude lower than for the 40 MeV neutron induced fission, still it is advantageous because the electrons/$\gamma$-photons conversion efficiency is much more significant than that for the deuterons/neutrons. In the photofission method for the neutron-rich radioactive ion beam production, nuclei are excited by photons covering the peak of the GDR where the energetic beam of incident electrons of $\sim$50 MeV can be slowed down in a tungsten (W) converter or directly in the target (U) itself, generating Bremsstrahlung $\gamma$-rays \cite{Es80} which can induce fission.  

    The aim of the present work is to obtain the photonuclear reaction cross sections at energies covering the GDR energy region in a framework of a two step process of the nuclear photoabsorption followed by the process of the competition between the light particle evaporation and fission for the excited nucleus. This model was used previously for the photonuclear reactions in the quasi deuteron (QD) \cite{Le51} energy region successfully \cite{Mu07,Mu09} whereas in the present work we explore the nuclei excited by photons in the GDR energy region which is particularly important in relation to the production of neutron rich nuclei. 

\noindent
\section{The GDR photoabsorption, the nuclear excitation and fission }
\label{section2}

    In the total photoabsorption cross section $\sigma_a^T$ at energies covering the GDR region, both the Lorentz type GDR cross section $\sigma_a^{GDR}$ and the QD cross section $\sigma_a^{QD}$ (albeit small) contribute and therefore 

\begin{equation}
 \sigma_a^T=\sigma_a^{GDR}+\sigma_a^{QD}. 
\label{seqn1}
\end{equation}
\noindent

    In the hydrodynamic theory of photonuclear reactions, the giant dipole resonance consists of Lorentz line for spherical nuclei \cite{Go48,St50}, corresponding to the absorption of photons which induce oscillations
of the neutron and proton fluids in the nucleus against each other, and the superposition of two such lines
for statically deformed spheroidal nuclei \cite{Ok56,Da58}, corresponding to oscillations along each of the nondegenerate axes of the spheroid. The lower-energy line corresponds to oscillations along the longer axis and the higher-energy line along the shorter, since the absorption frequency decreases with increasing nuclear dimensions. Therefore, the semiclassical theory of the interaction between photons and nuclei entails that the shape of a fundamental resonance in the absorption cross section is given by \cite{St50,Da58} 

\vspace{-0.0cm}
\begin{equation}
\sigma_a^{GDR}(E_{\gamma})= {\bf \huge\Sigma}_{i=1}^2 \frac{\sigma_i}{1+{\Big [} \frac{(E_{\gamma}^2-E_i^2)}{E_{\gamma} \Gamma_i}{\Big ]} ^2}
\label{seqn2}
\vspace{-0.0cm}
\end{equation}
\noindent
where $\sigma_i$, $E_i$ and $\Gamma_i$ are the peak cross section, resonance energy and full width at half maximum, respectively.

\begin{figure}[htbp]
\vspace{0.0cm}
\eject\centerline{\epsfig{file=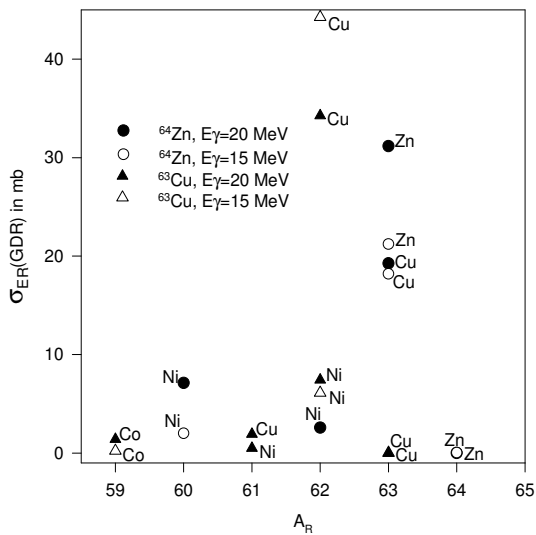,height=7cm,width=7cm}}
\caption
{The plots of cross sections $\sigma_{ER}^{GDR}$ as a function of mass number $A_R$ of the evaporation residues for $^{63}$Cu and $^{64}$Zn at $E_{\gamma}$=15 MeV, 20 MeV. For medium mass nuclei such as  $^{63}$Cu or $^{64}$Zn, there are no fission events at these energies.}
\label{fig1}
\vspace{1.4cm}
\end{figure}
\noindent

\begin{table}[h]
\vspace{-0.0cm}
\caption{\label{tab:table1} Comparison of the calculated photofission cross sections for GDR and QD contributions. }
\begin{tabular}{|c|c|c|c|c|c|c|}
\hline
Target&Fissility&$E_{\gamma}$&$\sigma_a^{GDR}$  &$\sigma_f^{GDR}$&$\sigma_a^{QD}$ &$\sigma_f^{QD}$\\
nuclei&$\frac{Z^2}{A}$& {\scriptsize MeV} & mb & mb & mb & mb          \\  
\hline
$^{237}$Np&36.49 &10&244.0&237.0&1.165&1.131\\ 
&&15&392.6&391.7&5.442&5.427\\ 
&&20&92.5&91.9&10.306&10.235\\
\hline
$^{238}$U &35.56&10  &269.1&186.4&1.140&0.787\\ 
&&15 &360.0&336.7&5.361&5.011\\ 
&&20 &73.7&64.9&10.189&8.967\\ 
\hline
$^{232}$Th&34.91&10     &247.5&17.7 &1.259&0.092\\ 
&&15&355.6&66.5&5.680&1.080\\
&&20&75.3&15.7&10.574&2.212\\  
\hline
\end{tabular} 
\vspace{-0.0cm}
\end{table}

    The QD model \cite{Le51} is based on the assumption that incident photon is absorbed by a correlated n-p pair inside the nucleus, leaving the remaining nucleons as spectators and is proportional to available number of n-p pairs inside nucleus and to the free deuteron photodisintegration cross section $\sigma_d(E_{\gamma})$. Thus 

\begin{equation}
 \sigma_a^{QD}(E_{\gamma}) = \frac{L}{A} NZ \sigma_d(E_{\gamma}) e^{-D/E_{\gamma}}
\label{seqn3}
\end{equation}
\noindent
where $N$, $Z$ and $A$ are the neutron, proton and mass numbers respectively, $L/A$ factor represents the fraction of correlated n-p pairs and the function $e^{-D/E_{\gamma}}$ accounts for the reduction of the n-p phase space due to the Pauli exclusion principle. A systematic study of total nuclear photoabsorption cross section data in the intermediate energy range shows that $D = 0.72 A ^{0.81}$ MeV \cite{Te89} which for photon energies upto the pion threshold, agrees reasonably well with the approach based upon phase space considerations \cite{Ch91} using Fermi gas state densities that conserve linear momentum for the Pauli blocking effects in the quasideuteron regime of hard photon absorption. The free deuteron photodisintegration cross section is given by $\sigma_d(E_{\gamma})=\frac{61.2[E_{\gamma} - B]^{3/2}}{E_{\gamma}^3}$ mb \cite{Wu77} where $B$=2.224 MeV is the binding energy of the deuteron. The quasideuteron model \cite{Le51} of nuclear photoabsorption is used together with modern rms radius data to obtain Levinger's constant $L = 6.8 - 11.2 A^{-2/3} + 5.7 A^{-4/3}$ of nuclei throughout the Periodic Table and is in good agreement \cite{Ta92} with those obtained from experimentally measured $\sigma_a^{QD}$ values. 

    The recoiling nucleus can be viewed as a compound nucleus having the same composition as the target nucleus but with an excitation energy of $E^*$ given by $E^*=m_0c^2$[(1+2$E_{\gamma}/m_0c^2)^{1/2}-$1] \cite{Ba09} where $E_{\gamma}$ is the photon energy and $m_0$ is the rest mass of the nucleus before photon absorption. This excited compound nucleus then undergoes successive evaporation of neutrons and light particles or fission. Hence the photonuclear reaction cross section $\sigma_r^X$ is a product of the nuclear photoabsorption cross section $\sigma_a^X$ and the statistical decay probability $\frac{\Gamma_r}{\Gamma}$ and is, therefore, given by $\sigma_r^X=\sigma_a^X.\frac{\Gamma_r}{\Gamma}$ where $\Gamma_r$ and $\Gamma$ are the partial and the total reaction widths respectively, and depending upon the case $X$ stands for the $GDR$ or the $QD$ contributions. 

\noindent
\section{Competition between light particle evaporation and fission}
\label{section3}

    The statistical approach for nucleon and light-particle evaporation and nuclear fission is an appropriate scheme for calculation of the relative probabilities of different decay modes of the compound nucleus. Such statistical decay of the compound nucleus is the slow stage of the photonuclear reaction. The full Hauser-Feshbach formalism is used in the present calculations, however, for pedagogical reasons, we outline the formal calculations in the Weisskopf limit only. According to the standard Weisskopf evaporation scheme \cite{We37}, the partial width $\Gamma_j$ for the evaporation of a particle $j$ = n, p, $^2$H, $^3$H, $^3$He or $^4$He is given by

\begin{equation}
 \Gamma_j = \frac{(2s_j+1)\mu_j}{\pi^2\hbar^2\rho_{CN}(E^*)} \int_{V_j}^{E^*-B_j} \sigma^j_{inv}(E)\rho_j(E^*-B_j-E) EdE
\label{seqn4}
\end{equation}
\noindent
where $s_j$, $\mu_j$, $V_j$ and $B_j$ are the spin, reduced mass, Coulomb barrier and the binding energy of the particle $j$, respectively. $\sigma^j_{inv}(E)$ is the cross section for the inverse reaction which means the capture reaction cross section of the particle $j$ to create the compound nucleus. $\rho_{CN}$ and $\rho_j$ are the nuclear level densities for the initial and final (after the emission of the particle $j$) nuclei, respectively.

    The Bohr-Wheeler statistical approach \cite{Bo39} is used to calculate the fission width of the excited compound nucleus. This width is proportional to the nuclear level density $\rho_f$ at the fission saddle point:

\begin{equation}
 \Gamma_f = \frac{1}{2\pi\rho_{CN}(E^*)} \int_{0}^{E^*-B_f} \rho_f(E^*-B_f-E) dE
\label{seqn5}
\end{equation}
\noindent
where $B_f$ is the fission barrier height. The diffuse surface nucleus Sierk's \cite{Si79} fission barriers ($B_f$) are used for these calculations. The decay of the excited compound nucleus is treated in full Hauser-Feshbach formalism \cite{Ha52} and simulated using the Monte-Carlo method \cite{Pa84}. Finally, in order to calculate the fission probability, the total number of fission events in a computer run is counted and divided by the total number of photoabsorption events. Evaporation from excited fission fragments is also taken into account.

    The quantitative description of the process is based on the liquid drop model (LDM) for nuclear fission by Bohr and Wheeler \cite{Bo39} and the statistical model of nuclear evaporation originally developed by Weisskopf \cite{We37}. Accordingly, the probability of fission relative to neutron emission in the Weisskopf's limit is given by the Vandenbosch-Huizenga's equation \cite{Hu73}

\begin{equation}
 \frac{\Gamma_f}{\Gamma_n} = \frac{K_0 a_n[2(a_f E_f^*)^\frac{1}{2}-1]}{4A^\frac{2}{3}a_f E_n^*} \exp{{\big [}2 [(a_f E_f^*)^\frac{1}{2} - (a_n E_n^*)^\frac{1}{2}] {\big ]}}
\label{seqn6}
\end{equation}
\noindent
where $E_n^*=E^*-B_n$ and $E_f^*=E^*-B_f$ are the nuclear excitation energies after the emission of a neutron and after fission, respectively, where $B_n$ is the binding energy of the emitted neutron. $\Gamma_n$ and $\Gamma_f$ are the partial widths for the decay of the excited compound nucleus via neutron emission and fission, respectively, and the parameters $a_n$ and $a_f$ are the level density parameters for the neutron emission and the fission, respectively and $K_0=\hbar^2/2mr^2_0$ where $m$ and $r_0$ are the neutron mass and radius parameter respectively. The emission probability of particle $k$ relative to neutron emission according to the Weisskopf's statistical model \cite{We37} is given by

\begin{equation}
 \frac{\Gamma_k}{\Gamma_n} = {\Big (}\frac{\gamma_k}{\gamma_n}{\Big )} {\Big (}\frac{E_k^*}{E_n^*}{\Big )}  {\Big (}\frac{a_k}{a_n}{\Big )} \exp{{\big [}2 [(a_k E_k^*)^\frac{1}{2} - (a_n E_n^*)^\frac{1}{2}] {\big ]}}
\label{seqn7}
\end{equation}
\noindent
where $a_k$ is the level density parameter for the emission of the particle $k$, $\gamma_k/\gamma_n =1$ for $k=p$, 2 for $k=^4$He, 1 for $k=^2$H, 3 for $k=^3$H and 2 for $k=~^3$He. $E_k^* = E^* - (B_k + V_k)$ is the nuclear excitation energy after the emission of particle $k$ \cite{Le50}. $B_k$ are the binding energies of the emitted particles and $V_k$ are the Coulomb potentials. 

    The evaporation-fission competition of the compound nucleus is then described in the framework of a projection angular momentum coupled \cite{Ha52} Monte-Carlo routine \cite{Pa84}. It follows the correct procedure of angular momentum coupling at each stage of de-excitation and considers statistical emissions of $\gamma$, n, p and $\alpha$ particles. Any particular reaction channel $r$ is then defined as the formation of the compound nucleus via photoabsorption and its decay via particle emission or fission. Thus, fission is considered as a decay mode. The transmission coefficients for light-particle emissions are determined from the optical model potentials. The shift in the coulomb barrier during de-excitation is accounted by calculating the transmission coefficients at an effective energy determined by the shift. The Gilbert-Cameron level density \cite{GC65} is adopted in the calculations. The photonuclear reaction cross sections $\sigma_r^X$ are calculated using the equation $\sigma_r^X = \sigma_a^X n_r/N$ where $n_r$ is the number of events in a particular reaction channel $r$ and $N$ is total number of events that is the number of the incident photons. 

\begin{table*}[htbp]
\vspace{-0.0cm}
\caption{\label{tab:table1} Variation of the calculated photofission cross sections of actinides with incident photon energy $E_{\gamma}$.}
\begin{tabular}{cccccccccccccc}
\hline
\hline
Target&Cross&$E_{\gamma}$=&8.0&9.0&10.0 &11.0 &12.0 &13.0 &14.0 &15.0 &16.0&17.0&18.0 \\
nuclei&section&&{\scriptsize MeV}&{\scriptsize MeV}&{\scriptsize MeV}&{\scriptsize MeV}&{\scriptsize MeV}&{\scriptsize MeV}&{\scriptsize MeV}&{\scriptsize MeV}&{\scriptsize MeV}&{\scriptsize MeV}&{\scriptsize MeV}\\ \hline
\hline
$^{237}$Np&$\sigma_a^{GDR}$(Calc.)&mb&70.50&128.86 &244.01 &374.63 &390.37 &414.86 &443.94 &392.63 &296.09&213.54 &156.45\\
                 &$\sigma_f^{GDR}$(Calc.)&mb&69.70&126.72 &236.82 &354.73 &373.54 &399.18 &441.63 &391.75 &295.39&212.51 &155.20\\ 
&$\sigma_a^{GDR}$(Expt.)&mb&&122&234&372&391&409&441&391&297&&\\
&$\sigma_f^{GDR}$(Expt.)&mb&&66&134&219&197&225&284&259&178&&\\ \hline
                
$^{238}$U&$\sigma_a^{GDR}$(Calc.)&mb&69.39&133.20 &269.12 &411.46 &393.40 &412.75 &435.88 &360.02 &253.74 &175.90 &126.37\\
                 &$\sigma_f^{GDR}$(Calc.)&mb&50.76&87.61  &186.55 &257.68 &241.51 &329.80 &410.53 &336.27 &224.60 &154.63 &104.90\\
&$\sigma_a^{GDR}$(Expt.)&mb&69&125&266&406&391&409&431&359&262&172&125\\
&$\sigma_f^{GDR}$(Expt.)&mb&23&31&59&91&87&119&153&134&100&69&53\\ \hline
   
$^{232}$Th&$\sigma_a^{GDR}$(Calc.)&mb&71.28&131.69 &247.51 &373.36 &389.29 &409.94 &427.28 &355.55 &253.63 &177.37 &128.18 \\
                 &$\sigma_f^{GDR}$(Calc.)&mb&0.39&10.28   &17.99   &34.46   &29.48   &54.65   &132.97 &66.58  &48.12   &34.08 &22.41\\ 
&$\sigma_a^{GDR}$(Expt.)&mb&&144&266&375&394&412&431&350&250&&\\
&$\sigma_f^{GDR}$(Expt.)&mb&&12&16&31&16&31&47&37&31&&\\ \hline
\hline
\end{tabular} 
\vspace{-0.0cm}
\end{table*}

\noindent
\section{ Calculation and results }
\label{section4}

    Each calculation is performed with 40000 events using a Monte-Carlo technique for the evaporation-fission calculation. This provides a reasonably good computational statistics. The photonuclear reaction cross sections are calculated for the GDR and the QD contributions at different energies for various elements. The Lorentz line parameters $\sigma_1,E_1,\Gamma_1$ and $\sigma_2,E_2,\Gamma_2$, for the GDR cross sections are taken from Refs.\cite{Ve73,Di88} and their mean values are used for the present calculations. In Table-1, the results of these calculations for the GDR and the QD contributions for three actinide nuclei are listed and compared. In Fig. 1, the cross sections $\sigma_{ER}^{GDR}$ as a function of mass number $A_R$ of the evaporation residues are plotted for $^{63}$Cu and $^{64}$Zn at $E_{\gamma}$=15 MeV, 20 MeV. For medium mass nuclei such as $^{63}$Cu or $^{64}$Zn, there are no fission events \cite{Ro03} at these energies. The variation of $\sigma_a^{GDR}$ and $\sigma_f^{GDR}$ at many gamma-ray energy points over the entire GDR bump are detailed in Table-2 and compared with the corresponding experimental values which are extracted from the plots of Ref.\cite{Ve73}. Although the calculated $\sigma_a^{GDR}$ values agree quite well with the experimental values, the photofission cross sections $\sigma_f^{GDR}$ are overestimated implying more neutron evaporation than estimated by the calculations. The statistical error and standard errors in the Lorentz line parameters \cite{Ve73} can not account for the uniform overestimation. On the contrary, at higher energies the situation appears to be reversed and the present calculations slightly overestimated the neutron multiplicities \cite{Mu09}. However, the present results corroborate  our earlier findings at higher energies  \cite{Mu07} where the fission probabilities for $^{232}$Th, $^{238}$U and $^{237}$Np were found to be as high as $\sim$20$\%$, $\sim$87$\%$ and $\sim$99$\%$ respectively. The main physics issue probably remains in the second stage of the process which suggests perhaps the need for more precise experimental measurements in the actinide region since little overcounting of the evaporation neutrons in the measurements can account for this discrepancy.

    The upper limits of the reaction cross sections can be estimated using $\sigma_f^X=\sigma_a^X$/N for cases where not a single fission event occurred in N events, where N=40000 is the number of incident photons. The statistical error $\Delta\sigma_r^X$ in the theoretical estimates for the photonuclear reaction cross sections $\sigma_r^X$ can also be calculated using the equation $\sigma_r^X \pm \Delta\sigma_r^X=\sigma_a^X [n_r \pm \sqrt{n_r}]/N$ which implies that the statistical error in the photonuclear reaction cross section $\Delta\sigma_r^X=\sqrt{\sigma_a^X\sigma_r^X/N}$. 

\noindent
\section{ Summary and conclusion }
\label{section5}

    In summary, the photonuclear reaction cross sections are calculated for photon induced reactions in the GDR energy region within a Monte-Carlo framework for simulation of the evaporation-fission competition. The calculations are performed assuming 40000 incident photons for each calculation which provide reasonably good statistics for computationally stable results. This model was used previously for the photonuclear reactions in the QD energy region successfully \cite{Mu07,Mu09} whereas in the present work we explore the nuclei excited by photons at the GDR energy region which is particularly important in relation to the production of neutron rich nuclei.  

    The present calculations provide excellent estimates of photoabsorption cross section for the actinides while the photofission cross sections are somewhat overestimated implying more neutron evaporation than expected. The light target nuclei such as $^{63}$Cu, $^{64}$Zn, are not very useful for RIB production. The contribution to the photofission cross sections from the QD process is very small compared to that from the GDR. 
The fission channel for the medium mass nuclear targets such as $^{63}$Cu and $^{64}$Zn, is analyzed and found to be negligible in our calculations.

\end{document}